\documentclass[12pt]{article}
\usepackage{amsmath}
\usepackage{graphicx,psfrag,epsf}
\usepackage{epsfig}
\usepackage{epstopdf}
\usepackage{enumerate}
\usepackage{natbib}
\usepackage{algpseudocode}
\usepackage{url} % not crucial - just used below for the URL
\usepackage{hyperref, bm}
\usepackage{amsmath,amssymb}
\usepackage{bbm}
\usepackage{xcolor}
\usepackage{caption}
\usepackage{enumitem}
\usepackage{array}
\usepackage{subcaption}
\usepackage{multirow}
\usepackage{amsfonts,amstext,amsmath,amssymb,amsthm}
\usepackage{mathtools}
\usepackage{mathabx} 
\usepackage{algorithm}
\newcolumntype{P}[1]{>{\centering\arraybackslash}p{#1}}

\newtheorem{theorem}{Theorem}

\newcommand{\argmin}{\mathop{\mathrm{argmin}}}
\newcommand{\argmax}{\mathop{\mathrm{argmax}}}
\DeclareMathOperator*{\arginf}{arg\,inf}

\newcommand{\blind}{0}
\begin{document}

\def\spacingset#1{\renewcommand{\baselinestretch}%
{#1}\small\normalsize} \spacingset{1}

%%%%%%%%%%%%%%%%%%%%%%%%%%%%%%%%%%%%%%%%%%%%%%%%%%%%%%%%%%%%%%%%%%%%%%%%%%%%%%

\if0\blind
{
  \title{\bf Multivariate Rank-Based Analysis of Multiple Endpoints in Clinical Trials: A Global Test Approach}
\author{
  Kexuan Li\thanks{kexuan.li.77@gmail.com, Global Analytics and Data Sciences, Biogen, Cambridge, Massachusetts, US.},
  Lingli Yang\thanks{lyang8@wpi.edu, Department of Mathematical Sciences, Worcester Polytechnic Institute, Massachusetts, US.},
  Shaofei Zhao\thanks{shaofei.zhao@abbvie.com, Data and Statistical Sciences, AbbVie, Madison, New Jersey, US.},\\
  Susie Sinks\thanks{susie.sinks@biogen.com, Global Analytics and Data Sciences, Biogen, Cambridge, Massachusetts, US.},
  Luan Lin\thanks{luan.lin@biogen.com, Global Analytics and Data Sciences, Biogen, Cambridge, Massachusetts, US.},
  Peng Sun\thanks{peng.sun@biogen.com, Global Analytics and Data Sciences, Biogen, Cambridge, Massachusetts, US.}
}
\maketitle
} \fi

\if1\blind
{
  \bigskip
  \bigskip
  \bigskip
  \begin{center}
    {\LARGE\bf Title}
\end{center}
  \medskip
} \fi

\bigskip
\begin{abstract}
Clinical trials often involve the assessment of multiple endpoints to comprehensively evaluate the efficacy and safety of interventions. In the work, we consider a global nonparametric testing procedure based on multivariate rank for the analysis of multiple endpoints in clinical trials. Unlike other existing approaches that rely on pairwise comparisons for each individual endpoint, the proposed method directly incorporates the multivariate ranks of the observations. By considering the joint ranking of all endpoints, the proposed approach provides robustness against diverse data distributions and censoring mechanisms commonly encountered in clinical trials. Through extensive simulations, we demonstrate the superior performance of the multivariate rank-based approach in controlling type I error and achieving higher power compared to existing rank-based methods. The simulations illustrate the advantages of leveraging multivariate ranks and highlight the robustness of the approach in various settings. The proposed method offers an effective tool for the analysis of multiple endpoints in clinical trials, enhancing the reliability and efficiency of outcome evaluations.
\end{abstract}

\noindent%

\spacingset{1.25}

\newpage
\section{Introduction}
In clinical trials, patients are often evaluated using multiple measures of treatment effectiveness, such as survival time, biomarker dynamics, and functional evaluations. To ensure a comprehensive evaluation of therapeutic benefits, it is important to consider multiple endpoints simultaneously. For instance, spinal muscular atrophy (SMA), a rare neuromuscular disorder causing motor neuron loss and progressive muscle degeneration, is assessed in treatment trials by comparing therapies based on both survival time and changes in muscle function measured using the Hammersmith Functional Motor Scale-Expanded (HFMSE). To draw a conclusive judgment on the effectiveness of treatment across multiple outcomes, several approaches have been suggested to summarize the treatment effect. One such approach involves employing multiple testing correction techniques, which adjust the individual $p$-values obtained from statistical tests to account for the possibility of false positives. Examples of such correction methods include the Bonferroni correction \citep{bonferroni1936teoria} and the Benjamini-Hochberg procedure \citep{benjamini1995controlling}. Nevertheless, these methods are unable to provide a comprehensive assessment regarding the overall efficacy of clinical intervention. Furthermore, the endpoints measured in a clinical trial often exhibit high correlation, which introduces further challenges when utilizing these approaches. Often, however, people might be more interested in highlighting whether a subset of variables is jointly suggestive of a treatment effect \citep{ristl2019methods}. In such cases, instead of testing each outcome individually, an overall statement should be claimed to assess therapeutic benefit. An alternative approach to address the issue of multiplicity is to create a composite endpoint by merging all relevant clinical outcomes into a single variable. This allows for an evaluation of the overall therapeutic benefit in a comprehensive manner, providing a ``global" assessment. By conducting a single statistical test on the composite endpoint, there is no need for adjustment or correction for multiple comparisons. This approach simplifies the analysis and interpretation process by considering all relevant outcomes simultaneously. In essence, the global test procedure transforms a multivariate problem into a univariate scale, enabling the announcement of a single probability statement regarding the success of targeted intervention strategies. By utilizing a global testing procedure, it becomes possible to summarize the overall effect of treatment across multiple outcomes in a more flexible manner. This approach allows for a comprehensive evaluation of treatment efficacy, considering the collective impact of various outcomes simultaneously. In this paper, we introduce a nonparametric global testing procedure based on (multivariate) rank energy statistics developed by \cite{deb2021multivariate}. This procedure is designed to summarize the overall treatment effect across multiple measurements, including censored outcomes. The proposed method enables the integration of various measurements, accommodating both continuous, discrete, and censored data, to provide a comprehensive assessment of the treatment's impact across multiple measurements, including censored outcomes.

Before proceeding, let us first review some of the global test procedures that have been proposed in the literature. To name a few, \cite{o1984procedures} proposed
a nonparametric rank-sum-type test to compare the distribution of two samples with multiple outcomes by summing up the ranks for each individual outcome and the test statistics is proven to be asymptotically normal distributed under the null hypothesis that the two multivariate samples have the same distribution. \cite{finkelstein1999combining} presented a nonparametric test for time to event endpoint and longitudinal measure. \cite{wittkowski2004combining} introduced a family of simple testing procedures for scoring multivariate ordinal data. \cite{huang2008rank} considered the sample size computation problem for clinical trial design with multiple primary outcomes. \cite{buyse2010generalized} constructed a hierarchical global ranking of a survival endpoint and a longitudinal measure to test the null hypothesis that neither of the outcomes being associated with treatment. \cite{berry2013combined} described a new endpoint for amyotrophic lateral sclerosis by combining survival time and change in function score. \cite{ramchandani2016global} further generalized the aforementioned global
nonparametric rank tests using U-statistics and discussed the choice of optimal
weighting schemes. Specifically, Let $x_{ik}, y_{jk}$ be the observed outcome $k$ for subject $i$ in the control group, $i=1, \ldots, m$  and subject $j$ in the treatment group, $j=1, \ldots, n$, where $k=1, \ldots, d$. Denote $\bm{x}_{i} = (x_{i1}, \ldots, x_{id})^\top, \bm{y}_{j} = (y_{j1}, \ldots, y_{jd})^\top$ as the observed vector. \cite{ramchandani2016global} defined $r_{ij}^{(k)}(x_{ik}, y_{jk}):\mathbb{R}\times\mathbb{R} \rightarrow[-1,1]$ be the rank score between $i$-th subject in control group and $j$-th subject in the treatment group for outcome $k$. For example, $r_{ij}^{(k)}(x_{ik}, y_{jk}) = \mathbbm{1}(x_{ik} > y_{jk}) - \mathbbm{1}(x_{ik} < y_{jk})$.  \cite{ramchandani2016global} further defined $\bm{r}_{ij}(\bm{x}_i, \bm{y}_j) = (r_{ij}^{(1)}(x_{i1}, y_{j1}),\ldots, r_{ij}^{(d)}(x_{id}, y_{jd}))$ as the vector of rank scores brtween $i$-th subject and $j$-th subject, and for simplicity, we sometimes write $\bm{r}_{ij}$ for $\bm{r}_{ij}(\bm{x}_i, \bm{y}_j)$, $r_{ij}^{(k)}$ for $r_{ij}^{(k)}(x_{ik}, y_{jk})$ if no confusion arises. Then, the two-sample U-statistics generalized by \cite{ramchandani2016global} was defined as the summation among all the pairwise comparisons between two groups after mapping $\bm{r}_{ij}$ to a univariate score:
\begin{equation}
U = \frac{1}{mn}\sum_{i=1}^m\sum_{j=1}^n\phi(\bm{r}_{ij}),
\end{equation}
where $\phi:\mathbb{R}^d\rightarrow\mathbb{R}$ maps the vector of comparison for each outcome to a one-dimensional score, which gives us the overall evaluation. Ideally, the maximizer and minimizer of $\phi(\bm{r}_{ij})$ should be $(1, \ldots, 1)$ and $(-1, \ldots, -1)$, respectively. Several choices of $\phi$ have been derived in literature. For example, in \cite{o1984procedures}, the author proposed a nonparametric procedure that calculated an overall rank by summing outcome-specific ranks for each subject and used a two-sample rank-sum or $t$ test to test for the null hypothesis. In other word, map $\phi$ in \cite{o1984procedures} is given by
    \begin{equation}
    \phi(\bm{r}_{ij}) = r_{ij}^{(1)} + r_{ij}^{(2)} +\ldots+r_{ij}^{(d)}.
    \end{equation}
In \cite{o1984procedures}, the underlying assumption is that each outcome has the same importance. However, in practice, not all the endpoints may contribute equally to the treatment effect, so, the rank-sum test by \cite{o1984procedures} could be further generalized to the weighted summation as $\phi(\bm{r}_{ij}) = \sum_{k=1}^pw_kr_{ij}^{(k)}$, where $w_k \geq 0$ is the weight associated with each component. The Finkelstein-Schoenfeld test introduced by \cite{finkelstein1999combining} compared a mortality outcome and a longitudinal outcome in a hierarchical manner, where subjects are first compared pairwise on survival using the Gehan scoring function \citep{gehan1965generalized}, and then on the longitudinal marker if it is indeterminate who survived longer. In their framework, the function $\phi$ is defined as 
\[
\phi(\bm{r}_{ij}) = r_{ij}^{(1)} + \mathbbm{1}(r_{ij}^{(1)} = 0)r_{ij}^{(2)} + \ldots + \mathbbm{1}(r_{ij}^{(1)} = \ldots = r_{ij}^{(d-1)} = 0)r_{ij}^{(d)}.
\]
In \cite{wittkowski2004combining}, the authors conducted pairwise comparisons of subjects based on ordinal measures. They considered two situations: (1) assigning a score of 1 to the subject whose outcomes are all favorable, and (2) assigning a score of 1 to the subject who has more favorable outcomes. The corresponding function $\phi$ can be expressed as 
\begin{align*}
\phi(\bm{r}_{ij}) &= \mathbbm{1}(\max_{k=1,\ldots, d} \{r_{ij}^{(k)}\} > 0) - \mathbbm{1}(\min_{k=1,\ldots, d} \{r_{ij}^{(k)}\} < 0), \text{ for situation (1)}\\
\phi(\bm{r}_{ij}) &= \mathbbm{1}(\sum_{k=1}^d r_{ij}^{(k)} > 0) - \mathbbm{1}(\sum_{k=1}^d r_{ij}^{(k)} < 0), \text{ for situation (2)}.
\end{align*}

All the methods mentioned above utilized the univariate rank of each individual endpoint and mapped each pair to a one-dimensional score. In addition to rank-based testing procedures, various other testing procedures have been explored in the literature for handling multiple endpoints, such as \cite{pocock2012win, luo2015alternative, dong2016generalized}. For a comprehensive review of these and other testing procedures, please refer to \cite{ristl2019methods}.

The remainder of the paper is organized as follows. Section \ref{Method} formulates the statistical problem and presents nonparametric global rank testing for survival
and multiple endpoints using multivariate ranks as well as the implementation. Section \ref{Simulation} evaluates the finite-sample performance of the proposed methodology by several simulation studies and some concluding remarks are provided in Section \ref{Conclusion}.

\section{Methodology} \label{Method}
Before presenting the proposed nonparametric global test statistics, we first formulate the problem under study. Suppose in a clinical trial, people collect $d$ measurements to verify the efficacy of an intervention. Let $x_{ik}, y_{jk}$ be the observed outcome $k$ for subject $i$ in the control group, $i=1, \ldots, m$  and subject $j$ in the treatment group, $j=1, \ldots, n$, where $k=1, \ldots, d$. Denote $\bm{x}_{i} = (x_{i1}, \ldots, x_{id})^\top, \bm{y}_{j} = (y_{j1}, \ldots, y_{jd})^\top$ as the observed vector. Suppose $\bm{x}_1, \ldots, \bm{x}_m \overset{\text{i.i.d.}}{\sim} {\mu}_{\bm{x}}$ and $\bm{y}_1, \ldots, \bm{y}_n \overset{\text{i.i.d.}}{\sim} {\mu}_{\bm{y}}$, where ${\mu}_{\bm{x}}, {\mu}_{\bm{y}}$ are two $d$-dimensional distributions. The following nonparametric two-sample goodness-of-fit testing problem is considered
\begin{eqnarray}\label{eq:testing_problem}
H_0: \mu_{\bm{x}} = \mu_{\bm{y}}, \,\,\, \textrm{versus} \,\,\, H_1: \mu_{\bm{x}} \neq \mu_{\bm{y}}.
\end{eqnarray}
When $d=1$ and ${\mu}_{\bm{x}}, {\mu}_{\bm{y}}$ are unknown, the problem is a classical nonparametric two-sample test and several methods have been proposed for it, which includes Spearman's rank correlation test \citep{spearman1961proof}, two-sample Cram\'er-von Mises statistic \citep{cramer1928composition}, two-sample Kolmogorov-Smirnov test \citep{smirnov1939estimation}, Wald-Wolfowitz run test \citep{wald1940test}, Wilcoxon-Mann-Whitney rank test \citep{mann1947test}, Hoeffding’s D-test \citep{hoeffding1994non}, among others. In the case when $d>1$, the problem of nonparametric two-sample testing for multivariate distributions has also a long history and has recently gained significant attention. Various methods have been proposed, such as those by \cite{bickel1969distribution, friedman1979multivariate, maa1996reducing, gretton2012kernel, chen2017new, mukhopadhyay2020nonparametric, liu2020learning, liu2021meta}, \cite{oja2004multivariate}, \cite{rosenbaum2005exact}, \cite{biswas2014distribution}. In this work, we specifically focus on the (multivariate) rank-based approach proposed by \cite{deb2021multivariate}.

%\subsection{Review of Nonparametric Two-sample Test for Multivariate Data}
%The nonparametric two-sample test for multivariate data has a long history and rich literature, such as, kernel-based approaches: graph-based approaches, machine-learning based approaches:\cite{bickel1969distribution, friedman1979multivariate,maa1996reducing, gretton2012kernel, chen2017new, mukhopadhyay2020nonparametric, liu2020learning, liu2021meta}. In this subsection, we focus on the rank-based approaches. \KL{\cite{oja2004multivariate}} rank-based: \cite{rosenbaum2005exact} proposed a distribution-free test statistics based on
%the minimum distance non-bipartite matching over the aggregate data and the number of pairs containing a sample from each group. \cite{biswas2014distribution} generalized the univariate Wald-Wolfowitz run test to high-dimensional data based on the
%shortest Hamiltonian path.

\subsection{Multivariate Rank}
\cite{deb2021multivariate} introduced the concept of multivariate rank, which utilizes low-discrepancy sequences to transform the original data into a unit hypercube. Without loss of generality, we let $\mathbb{R}^d$ be the $d$-dimensional input space, and the $d$-dimensional unit hypercube to which the data is mapped by the multivariate rank process is represented by $[0, 1]^d$. The families of all probability distributions on $\mathbb{R}^d$ are denoted by $\mathcal{P}(\mathbb{R}^d)$, while the families of Lebesgue absolutely continuous probability measures on $\mathbb{R}^d$ are represented by $\mathcal{P}_{ac}(\mathbb{R}^d)$, and the uniform distribution on $[0, 1]^d$ is denoted by $\mathcal{U}^d$. The multivariate rank approach is implemented using a measure transportation technique, also known as optimal transportation. Specifically, this involves finding a suitable function $G: \mathbb{R}^d \rightarrow \mathbb{R}^d$ that maps a given measure ${\mu} \in \mathcal{P}(\mathbb{R}^d)$ to ${\nu} \in \mathcal{P}(\mathbb{R}^d)$, represented as $G\#{\mu} = {\nu}$. In other words, if $\bm{x}$ follows the distribution ${\mu}$, then $G(\bm{x})$ follows the distribution ${\nu}$. Suppose the observed data follows distribution ${\mu} \in \mathcal{P}_{ac}(\mathbb{R}^d)$, then the idea of multivariate rank introduced by \cite{deb2021multivariate} is to find a rank function $R(\cdot)$ such that $R\#{\mu} = \mathcal{U}^d$, that is, $R(\bm{x})$ follows a uniform distribution in $\mathbb{R}^d$. The following theorem guarantees the existence of the (population) rank function $R(\cdot)$.
\begin{theorem}[McCann’s theorem \cite{mccann1995existence}]
Suppose ${\mu}, {\nu} \in \mathcal{P}_{ac}(\mathbb{R}^d)$, then there exists a unique  function $R(\cdot)$, up to measure zero sets with respect to ${\mu}$, which is the gradient of
an extended real-valued $d$-variate convex function, such that $R\#{\mu} = {\nu}$. Moreover, if ${\mu}$ and ${\nu}$ have finite second moments, then $R(\cdot)$ is also the solution to Monge’s problem, i.e.,
\begin{equation} \label{Monge}
R(\cdot) = \arginf_G\int\|\bm{x} - G(\bm{x})\|^2d{\mu}(\bm{x}), \,\,\,\textit{ subject to } G\#{\mu}={\nu}.    
\end{equation}
\end{theorem}
However, in practice, it is infeasible to know the true distribution ${\mu}$, instead, the only available information regarding the measure ${\mu}$ is obtained through empirical observations $\bm{x}_1, \ldots, \bm{x}_n$. \cite{deb2021multivariate} put forward a novel approach for estimating the (population) rank function using empirical observations. To understand this approach, it is first necessary to review the definition of the low-discrepancy sequence, which plays a crucial role in the method.

\begin{figure}[ht!]
  \centering
  \includegraphics[width=5in]{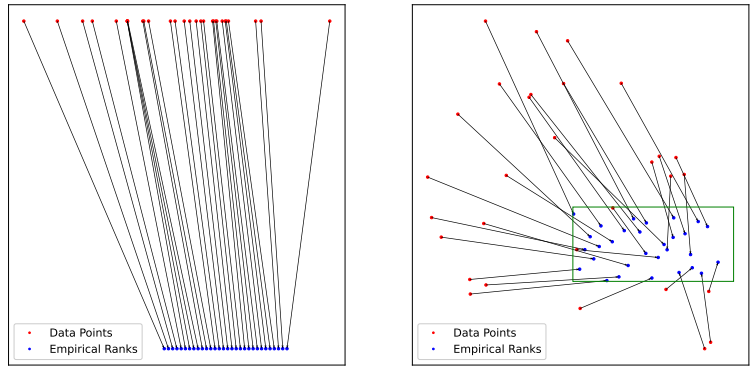}
  \caption{Left: Univariate rank on $[0, 1]$, the red dots are randomly generated from standard normal distribution, and the blue dots are evenly spaced on $[0, 1]$. The leftmost (smallest) red dot will be assigned to the first blue dot, thus rank 1, and so on. Right: Multivariate rank on $[0, 1]^2$, the blue dots are two-dimensional Sobol' sequence on $[0, 1]^2$, while the green box represents the region $[0, 1]^2$. The multivariate rank method will map the two-dimensional red dots to the blue dots, which correspond to their multivariate ranks.} \label{figure:multivariate_rank}
\end{figure}

A low-discrepancy sequence is a sequence of points in a multi-dimensional space that is designed to have better distribution properties than random sequences. In particular, these sequences are constructed to have a low discrepancy, which measures the uniformity of the distribution of points in a given region. To be more precise, let us consider a $d$-dimensional hypercube $[0,1]^d$ and let $\mathcal{A}$ be a set of $n$ points in this hypercube. The discrepancy of a set $\mathcal{A}$ is defined as:
\[
D(\mathcal{A}) = \sup_{\mathcal{B}\subset [0,1]^d} \left| \frac{\#(\mathcal{A}\cap \mathcal{B})}{n} - \text{Leb}(\mathcal{B}) \right|,
\]
where $\mathcal{B}$ is any region in $[0,1]^d$, $\#(\mathcal{A}\cap \mathcal{B})$ denotes the number of points in $\mathcal{A}$ that fall in $\mathcal{B}$, and $\text{Leb}(\mathcal{B})$ is the Lebesgue measure of $\mathcal{B}$. In other words, the discrepancy measures the maximum difference between the fraction of points in $\mathcal{A}$ that fall in any subinterval of $[0,1]^d$ and the measure of that subinterval. A low-discrepancy sequence is a sequence of points in $[0,1]^d$ that has a small discrepancy and the proportion of points in the sequence falling into an arbitrary set $\mathcal{B}$ is close to the proportional of the measure of $\mathcal{B}$. There are many different constructions of low-discrepancy sequences, such as Hammersley sequences \citep{hammersley1960monte}, Halton sequences \citep{Halton}, and Sobol sequences \citep{Sobol}. \citep{deb2021multivariate} proposed a novel approach for estimating the population rank function using a low-discrepancy sequence of points. In their method, they first generated a low-discrepancy sequence of points in the $d$-dimensional space of features, which was a good representation of $\mathcal{U}^d$. Then, the (empirical) rank map was defined as the solution of the empirical version of Monge's problem in (\ref{Monge}). To be more specific, let $\bm{x}_1,...,\bm{x}_n \in \mathbb{R}^d$ be the $i.i.d.$ observations, and $\{\bm{c}_1,...,\bm{c}_n\}\subset[0, 1]^d$ be a low-discrepancy sequence on $[0, 1]^d$. Let $\delta_{\bm{a}}$ denote the Dirac measure that assigns probability 1 to the point $\bm{a}$ and $\mu_n^{\bm{x}}(\bm{x}) = n^{-1}\sum_{i=1}^n\delta_{\bm{x}_i}, \nu_n = n^{-1}\sum_{i=1}^n\delta_{\bm{c}_i}$. Then the (empirical) rank map is defined as
\[
\widehat{R}(\cdot) = \arginf_F\int\|\bm{x} - F(\bm{x})\|^2d\mu_n^{\bm{x}}(\bm{x}), \,\,\,\textrm{ subject to } F\#\mu_n^{\bm{x}}=\nu_n,    
\]
which is equivalent to the following optimization problem:
\begin{equation}\label{assign}
      \widehat \sigma = \argmin_{\sigma=(\sigma(1),...,\sigma(n))\in S_n}\sum_{i=1}^n\parallel \bm{x}_i - \bm{c}_{\sigma(i)}\parallel^2  = \argmax_{\sigma=(\sigma(1),...,\sigma(n))\in S_n}\sum_{i=1}^n \langle\bm{x}_i,  \bm{c}_{\sigma(i)}\rangle,
\end{equation}
where $\parallel\cdot\parallel$ and $\langle \cdot, \cdot \rangle$ denote the usual Euclidean norm and inner product, and $S_n$ is the set of all permutations of $\{1, 2,..., n\}$. Finally, the (empirical) multivariate rank of $\bm{x}_i$ is a $d$-dimensional vector defined as
\begin{align}\label{rank}
\widehat{R}(\bm{x}_i) = \bm{c}_{\widehat \sigma(i)} ,\,\, \textrm{ for $i=1,\ldots,m$.}
\end{align}
Figure \ref{figure:multivariate_rank} illustrates the idea of univariate rank on $[0, 1]$ and multivariate rank on $[0, 1]^2$. In the following sections. we proceed to show how to construct a nonparametric global test for different types of endpoits based on multivariate rank.

\begin{figure}[ht!]
  \centering
  \includegraphics[width=5in]{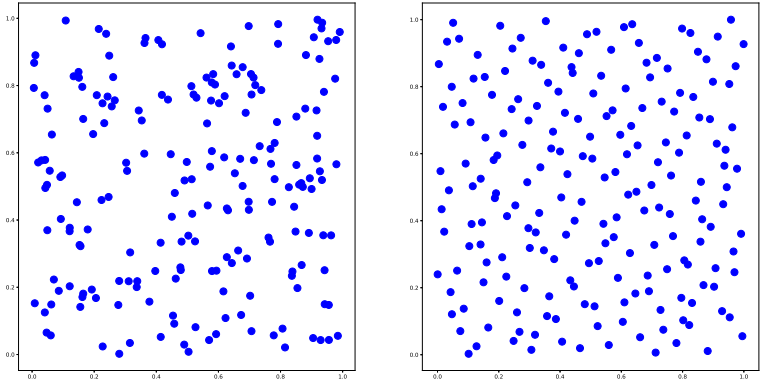}
  \caption{Left: The dots are randomly generated from a uniform distribution.
Right: The dots are generated using the Sobol' sequence. The Sobol' sequence provides a more even distribution of points compared to randomly generated points.} \label{figure:sobol_seq}
\end{figure}

\subsection{A Global Test Statistics Based on Multivariate Rank}
We now proceed to study the nonparametric global test problem in (\ref{eq:testing_problem}) based on multivariate rank. Given the observations of control group $\bm{x}_{1}, \ldots, \bm{x}_{m}$ and treatment group $\bm{y}_{1}, \ldots, \bm{y}_{n}$, first, pool the sample of $(m+n)$ observations into a single group and get the (empirical) multivariate rank of $\bm{x}_i, \bm{y}_j$ through (\ref{rank}), denoted as  $\widehat{R}_{m, n}^{\bm{x}, \bm{y}}\left(\bm{x}_i\right), \widehat{R}_{m, n}^{\bm{x}, \bm{y}}\left(\bm{y}_j\right)$. The corresponding test statistic based on multivariate rank is defined as
\begin{align}\label{eq:teststatitics}
\mathrm{RE}_{m, n}^2:=\frac{2}{m n} \sum_{i=1}^m & \sum_{j=1}^n\left\|\widehat{R}_{m, n}^{\bm{x}, \bm{y}}\left(\bm{x}_i\right)-\widehat{R}_{m, n}^{\bm{x}, \bm{y}}\left(\bm{y}_j\right)\right\|-\frac{1}{m^2} \sum_{i, j=1}^m\left\|\widehat{R}_{m, n}^{\bm{x}, \bm{y}}\left(\bm{x}_i\right)-\widehat{R}_{m, n}^{\bm{x}, \bm{y}}\left(\bm{x}_j\right)\right\| \nonumber \\
& -\frac{1}{n^2} \sum_{i, j=1}^n\left\|\widehat{R}_{m, n}^{\bm{x}, \bm{y}}\left(\bm{y}_i\right)-\widehat{R}_{m, n}^{\bm{y}, \bm{y}}\left(\bm{y}_j\right)\right\|.
\end{align}

Under Theorem 4.3 in \cite{deb2021multivariate}, we know $\mathrm{RE}_{m, n}^2$ is $O_p(1)$ with the following limiting distribution
\begin{align}\label{eq:limitingdistribution}
\frac{mn}{m+n}\mathrm{RE}_{m, n}^2 \xrightarrow{w} \sum_{j=1}^\infty\tau_jZ_j^2, \,\,\,\, \min(m,n) \rightarrow \infty,
\end{align}
where $Z_j$'s are $i.i.d.$ standard normals and $\tau_j$'s are fixed nonnegative constant which do not depend on the distribution $\bm{x}_i, \bm{y}_j$, and $\xrightarrow{w}$ denotes the weak convergence of distributions. Given predetermined significance level $\alpha$, let
\begin{equation}\label{eq:c_mn}
c_{m, n}:=\inf \left\{c>0: \mathbb{P}_{\mathrm{H}_0}\left(m n(m+n)^{-1} \mathrm{RE}_{m, n}^2 \geq c\right) \leq \alpha\right\},
\end{equation}
and the decision rule for testing (\ref{eq:testing_problem}) at significance level $\alpha$ can be defined as follows
\begin{equation*}
\phi_{m,n} = \mathbbm{1}(mn(m+n)^{-1}\mathrm{RE}_{m, n}^2 \geq c_{m, n}),
\end{equation*}
where $\mathrm{RE}_{m, n}^2, c_{m, n}$ are defined in (\ref{eq:teststatitics}) and (\ref{eq:c_mn}), respectively. We reject the null hypothesis in (\ref{eq:testing_problem}) if and only if $\phi_{m,n}=1$. By the definition of $c_{m, n}$, clearly, the test is level $\alpha$. It is worth noting that for any fixed $m, n$, $\mathrm{RE}_{m, n}^2, c_{m, n}$ do not depend on ${\mu}_{\bm{x}}, {\mu}_{\bm{y}}$. Unfortunately, as mentioned in \cite{deb2021multivariate}, the theoretical value of $c_{m,n}$ is infeasible to achieve, the only way to get $c_{m,n}$ as of now is through numerical experiments. We summarize the asymptotic thresholds for $mn(m+n)^{-1}\mathrm{RE}_{m, n}^2$ in Table \ref{table:asymototic} and provide the procedure to estimate $c_{m,n}$ in Algorithm \ref{alg:cmn}.

\begin{table}
\center
\caption{Asymptotic thresholds for $mn(m+n)^{-1}\mathrm{RE}_{m, n}^2$ when $\alpha=0.05, 0.10, d\leq6$.The numbers are obtained through Algorithm \ref{alg:cmn}.}
\label{table:asymototic}
\begin{tabular}{|l|l|l|l|l|l|l|}
\hline
           & $d=1$ & $d=2$ & $d=3$ & $d=4$ & $d=5$ & $d=6$\\ \hline
$\alpha=0.05$ &  0.94  & 1.12    & 1.26    & 1.37    &  1.45   & 1.54    \\ \hline
$\alpha=0.10$ &  0.70   &  0.92   &  1.07   &  1.17   &   1.28  &  1.37      \\ \hline
\end{tabular}
\end{table}

% \begin{table}
% \center
% \caption{Asymptotic thresholds for $mn(m+n)^{-1}\mathrm{RE}_{m, n}^2$ when $\alpha=0.05, 0.10, d\leq8$. \KL{The numbers are obtained through Algorithm XX.}}
% \label{table:asymototic}
% \begin{tabular}{|l|l|l|l|l|l|l|l|l|}
% \hline
%            & $d=1$ & $d=2$ & $d=3$ & $d=4$ & $d=5$ & $d=6$ & $d=7$ & $d=8$ \\ \hline
% $\alpha=0.05$ &  0.94  & 1.12    & 1.26    & 1.37    &  1.45   & 1.54    &  1.61   & 1.67    \\ \hline
% $\alpha=0.10$ &  0.70   &  0.92   &  1.07   &  1.17   &   1.28  &  1.37   &     &     \\ \hline
% \end{tabular}
% \end{table}

\begin{algorithm}
\caption{Algorithm of computing test
statistic $\mathrm{RE}_{m, n}^2$}\label{alg:test_statistics}
\begin{algorithmic}
\State \textbf{Input:}  $(\bm{x}_{1}, \ldots, \bm{x}_{m})$ and $(\bm{y}_{1}, \ldots, \bm{y}_{n})$.
\State Pool $(\bm{x}_{1}, \ldots, \bm{x}_{m})$ and $(\bm{y}_{1}, \ldots, \bm{y}_{n})$  into a single group.
\State Generate a low-discrepancy sequence $\{\bm{c}_1,...,\bm{c}_{(m+n)}\}$ on $[0, 1]^d$ with size $(m+n)$. 
\State Solve the optimal assignment problem:\\ 
\begin{center}
    $\widehat \sigma = \argmin_{\sigma=\left(\sigma(1),...,\sigma(m+n)\right)\in S_{m+n}}\left(\sum_{i=1}^{m}\parallel \bm{x}_i - \bm{c}_{\sigma(i)}\parallel^2+ \sum_{j=m+1}^{m+n}\parallel \bm{y}_{j-m} - \bm{c}_{\sigma(j)}\parallel^2\right)$.
\end{center}
\State Compute the (empirical) multivariate rank of $\bm{x}_i, \bm{y}_j$ by\\
\begin{center}
    $\widehat{R}_{m, n}^{\bm{x}, \bm{y}}(\bm{x}_i) =\bm{c}_{\widehat \sigma(i)}, i=1,\ldots,m, \widehat{R}_{m, n}^{\bm{x}, \bm{y}}(\bm{y}_j) =\bm{c}_{\widehat \sigma(j+m)}, j=1,\ldots,n$.
\end{center}
\State \textbf{Return}: test statistic $\mathrm{RE}_{m, n}^2$ in (\ref{eq:teststatitics}).
\end{algorithmic}
\end{algorithm}

\begin{algorithm}
\caption{Algorithm of estimating $c_{m,n}$ for any given $m,n, p, \alpha$}\label{alg:cmn}
\begin{algorithmic}
\State \textbf{Initialization:} set $n_{\textrm{run}} := 10^{6}$ and $A = [0, \ldots, 0]$ as a zero array with length $n_{\textrm{run}}$;
\State \textbf{Input:} $m, n >0, \alpha, p$;
\For{$i$ in $1, \ldots, n_{\textrm{run}}$}
\State generate $\bm{x}_1, \ldots, \bm{x}_m, \bm{y}_1, \ldots, \bm{y}_n$ from standard $d$-dimensional multivariate normal distribution independently;
\State compute test statistics $\mathrm{RE}_{m, n}^2$ using Algorithm \ref{alg:test_statistics};
\State $A[i] \leftarrow \mathrm{RE}_{m, n}^2$;
\EndFor
\State \textbf{Return}: $c_{m,n} = (1-\alpha)$ quantile of $A$.
\end{algorithmic}
\end{algorithm}
\subsection{Time-to-Event Endpoint} \label{sec:survival}
% use the idea in \citep{gehan1965generalized}
In the following subsection, we consider the problem that one of the endpoint is the time-to-event outcome. Without loss of generality, we assume the first component of the $d$ measurements is the survival endpoint and the rest $d-1$ components are non-survival endpoints. we assume $x_{i1}, y_{j1}$ be the survival endpoint and $x_{i1} = \min\{T^x_{i}, C^x_{i}\}, y_{j1} = \min\{T^y_{j}, C^y_{j}\}, \delta^x_{i}=\mathbbm{1}(T^x_{i} \leq C^x_i), \delta^y_{j}=\mathbbm{1}(T^y_{j} \leq C^y_j)$, where $T^x_{i} (T^y_{j}), C^x_{i} (C^y_j), \delta^x_{i} (\delta^y_j)$ is the unknown survival time, censoring time, and the censoring indicator of subject $i (j)$ in control(treatment) group. In order to apply the global test based on multivariate rank for the survival endpoint, we use the idea from the Gehan–Wilcoxon test \citep{gehan1965generalized}, which is an extension of the classical Wilcoxon rank-sum test for comparing survival curves between two or more groups. More specifically, in the first step, we pool the two samples of $(m+n)$ survival times $(x_{11}, \ldots, x_{m1}, y_{11}, \ldots, y_{n1})$ into a single group $(t_1, \ldots, t_{m+n})$, and we use a superscript `$+$' to indicate that the corresponding observation is censored. Then we construct a score by comparing each individual with the remaining $(m+n-1)$ subjects based on the following rule:
\begin{equation} \label{eq:u_tilde}
\widetilde{u}_{i j}=\left\{\begin{array}{ccc}
+1 & \text { if } & t_i>t_j \text { or } t_i^{+} \geq t_j, \\
-1 & \text { if } & t_i<t_j \text { or } t_i \leq t_j^{+}, \\
0 & \text { otherwise }.
\end{array}\right.
\end{equation}
Then the importance score for each individual is defined as $u_i = \sum_{j=1}^{m+n}\widetilde{u}_{ij}$. In other words, $U_i$ represents the number of survival (or censored) times which are \emph{definitely} less than $t_i$ (or $t_i^+$)  minus the number of survival (or censored) times which are \emph{definitely} greater than $t_i$ (or $t_i^+$). Once we get the importance score for each individual, the next step is straightforward. We can just easily replace the original survival times $(x_{11}, \ldots, x_{m1}, y_{11}, \ldots, x_{n1}, )$ with $(u_1, \ldots, u_{m+n})$. We summarize the procedure in Algorithm \ref{alg:survival1}.
\begin{algorithm}
\caption{A Global Multiple Rank-Based Procedure for Time-to-Event Endpoint}\label{alg:survival1}
\begin{algorithmic}
\State \textbf{Input:}  $(\bm{x}_{1}, \ldots, \bm{x}_{m})$ and $(\bm{y}_{1}, \ldots, \bm{y}_{n})$.
\State Pool $(x_{11}, \ldots, {x}_{m1})$ and $({y}_{11}, \ldots, {y}_{n1})$  into a single group as $(t_1, \dots, t_{m+n})$.
\State Compute the importance score $u_i = \sum_{j=1}^{m+n}\widetilde{u}_{ij}$ by (\ref{eq:u_tilde}).
\State Replace  $(x_{11}, \ldots, {x}_{m1})$ and $(y_{11}, \ldots, {y}_{n1})$ with $(u_1, \ldots, u_m)$ and $(u_{m+1}, \ldots, u_{m+n})$, respectively.
\State Compute test statistic $\mathrm{RE}_{m, n}^2$ using Algorithm \ref{alg:test_statistics} with input $(\widetilde{\bm{x}}_{1}, \ldots, \widetilde{\bm{x}}_{m})$ and $(\widetilde{\bm{y}}_{1}, \ldots, \widetilde{\bm{y}}_{n})$, where  $\widetilde{\bm{x}}_{i} = (u_i, x_{i2}, \ldots, x_{id})^\top, \widetilde{\bm{y}}_{j} = (u_{m+j}, y_{j2}, \ldots, y_{jd})^\top$.
\State \textbf{Return}: test statistic $\mathrm{RE}_{m, n}^2$.
\end{algorithmic}
\end{algorithm}
%\subsubsection{Time-to-Event Endpoint without Truncated Endpoints}
In the first step, we utilize the generalized Wilcoxon pairwise comparisons proposed by Gehan \citep{gehan1965generalized} to calculate the relative rank of the survival term for each subject. It is important to note that there are other methods available for obtaining relative ranks, such as imputation-based approaches \citep{efron1967two} or inverse probability of censoring weighting approach \citep{buyse2010generalized}. For a more comprehensive review of these methods, please refer to \cite{deltuvaite2022generalized}.

\section{Simulation} \label{Simulation}
In this section, we assess the finite sample performance of the global multivariate rank-based approach and compare it with two other rank-based approaches: O'Brien's rank-sum procedure \cite{o1984procedures} and Wittkowski’s method \citep{wittkowski2004combining}. In Wittkowski’s method, the pairwise comparison is based on $\phi(\bm{r}_{ij}) = \mathbbm{1}(\sum_{k=1}^d r_{ij}^{(k)} > 0) - \mathbbm{1}(\sum_{k=1}^d r_{ij}^{(k)} < 0)$, where a score of 1 is assigned if subject 1 has more favorable outcomes than subject 2. 
\subsection{Multiple Uncensored Outcomes}
\begin{figure}[ht!]
  \centering
  %\fbox{\rule[-.5cm]{0cm}{4cm} \rule[-.5cm]{4cm}{0cm}}
  \includegraphics[width=4.5in]{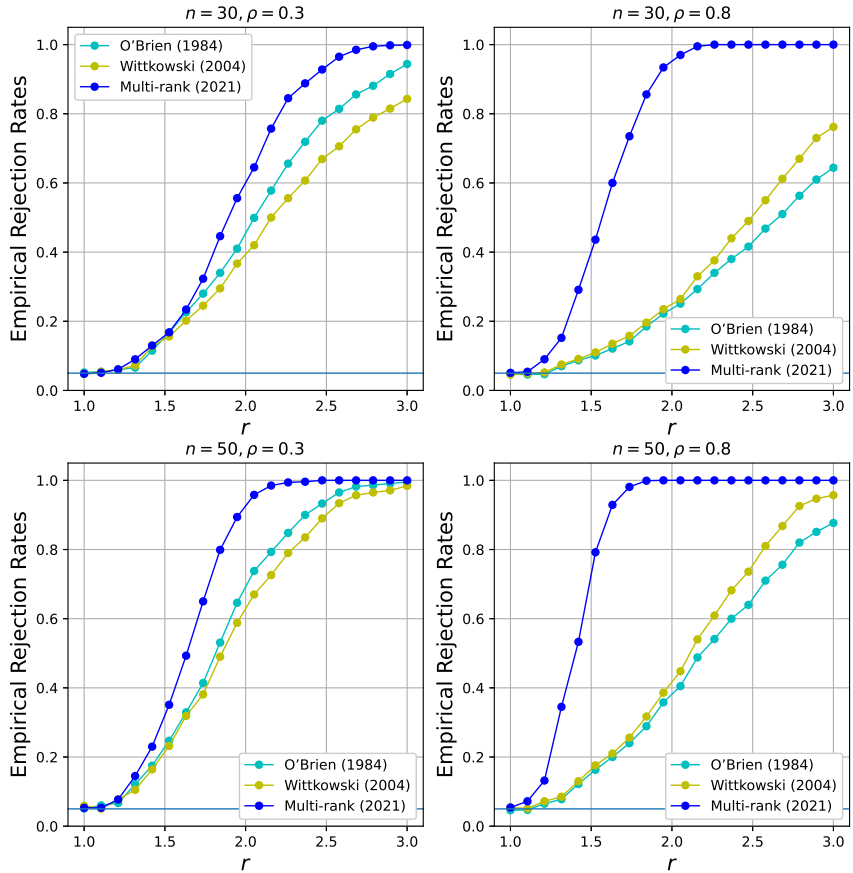}
  \caption{Empirical rejection rates for scenario 1.} \label{figure:scenario_1}
\end{figure}

\begin{figure}[ht!]
  \centering
 %\fbox{\rule[-.5cm]{0cm}{4cm} \rule[-.5cm]{4cm}{0cm}}
  \includegraphics[width=5in]{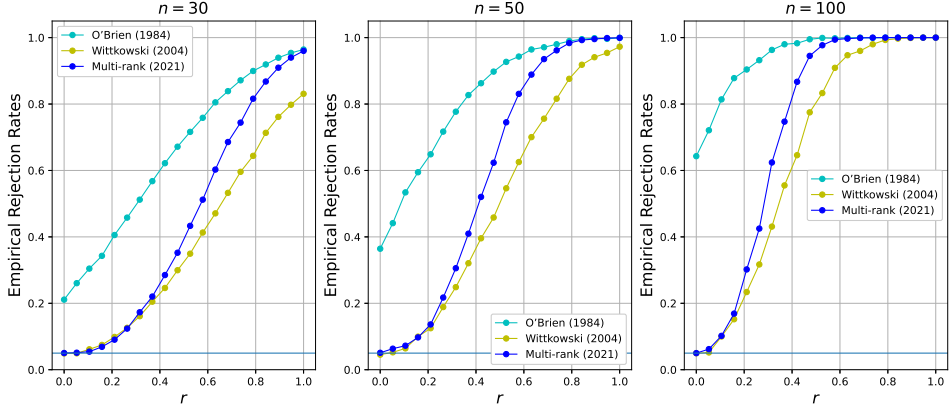}
  \caption{Empirical rejection rates for scenario 2.} \label{figure:scenario_2}
\end{figure}

In this section, we conduct simulation studies to evaluate the performance of the test procedure on uncensored endpoints. We consider two scenarios in which we examine both continuous and discrete endpoints.
\begin{itemize}[label={}]
\item \emph{scenario 1:} Suppose we collect $d=8$ measurements to verify the efficacy of an intervention and the observed values for two arms follow $\bm{x}_1, \ldots, \bm{x}_m \overset{\text{i.i.d.}}{\sim}\mathcal{N}(\bm{\mu}, \Sigma), \bm{y}_1, \ldots, \bm{y}_n \overset{\text{i.i.d.}}{\sim}\mathcal{N}(r\bm{\mu}, \Sigma)$, where $\bm{\mu}=(1, 0.1, 0.2, 0.3, 0.1, 0.8, 0.1, 0)^\top$  represents the mean vector, $\sigma_{ij} = 1$ if $i=j$, and $\sigma_{ij} = \rho$ if $i\neq j$ represents the $(i,j)$th entry of the covariance matrix $\Sigma$. We consider $\rho = 0.3, 0.8$ and $r$ from 1 to 3. In particular, $r=1$ is used to examine the empirical size of the proposed test under $H_0$, and other values of $r$ are used to check the empirical powers against alternatives. The target significance level is chosen as $\alpha = 0.05$. The results are summarized in Figure \ref{figure:scenario_1}. It can be observed that the type I error of all three methods is well controlled. However, the multivariate rank approach performs significantly better when $r>0$, indicating its superiority under $H_1$. An interesting observation is that Wittkowski's method performs better than O'Brien's method when the correlation between each endpoint is stronger.
\item \emph{scenario 2:} In this scenario, we consider four correlated endpoints ($d=4$) where three of them are continuous and one is discrete. Specifically, we let $\bm{x}_i = (x_{i1}, x_{i2}, x_{i3}, x_{i4})^\top, \bm{y}_j = (y_{j1}, y_{j2}, y_{j3}, y_{j4})^\top \in \mathbb{R}^4$, where
    \begin{align*}
    (x_{i1}, x_{i2}, x_{i3})^\top &\overset{\text{i.i.d.}}{\sim}\mathcal{N}(\bm{\mu}, \Sigma), \,\,\, (y_{j1}, y_{j2}, y_{j3})^\top \overset{\text{i.i.d.}}{\sim}\mathcal{N}(\bm{\mu} -r\bm{\nu}, \Sigma), \\
    x_{i4} &\overset{\text{i.i.d.}}{\sim} \text{Bernoulli}(p_{{x}}), \,\,\, y_{j4} \overset{\text{i.i.d.}}{\sim} \text{Bernoulli}(p_{{y}}),
    \end{align*}
    where 
    \begin{align*}
        p_{{x}} = \frac{\exp\{-3 + \beta_1x_{i1} + \beta_2x_{i2} + \beta_3x_{i3}\}}{1+\exp\{-3 + \beta_1x_{i1} + \beta_2x_{i2} + \beta_3x_{i3})\}},\,\,\, p_{{y}} = \frac{\exp\{-3 + \beta_1y_{j1} + \beta_2y_{j2} + \beta_3y_{j3}\}}{1+\exp\{-3 + \beta_1y_{j1} + \beta_2y_{j2} + \beta_3y_{j3}\}},
    \end{align*}
    $(\beta_1, \beta_2, \beta_3)^\top = (0.1, 0.4, 0.1)^\top, \bm{\mu} = (150, 6, 250)^\top, \bm{\nu} = (10, 0.1, 10)^\top, \Sigma = \begin{bmatrix}
10^2 & 7 & 0.6\\
7 & 1 & 0.4 \\
0.6 & 0.4 &15^2
\end{bmatrix}$.
For our analysis, we vary the treatment effect parameter $r$ from 0 to 1. Specifically, we set $r = 0$ to examine the empirical size of the proposed test under the null hypothesis $H_0$. The results are summarized in Figure \ref{figure:scenario_2}. It can be observed that when including a discrete endpoint, the type I error of the multivariate rank approach and Wittkowski's methods can still be well controlled at the target significance level. However, O'Brien's method shows an inflated type I error. Additionally, the multivariate rank approach exhibits greater power compared to Wittkowski's method.
\end{itemize}

\subsection{Time-to-Event Endpoint} \label{sec:sim_survival}
\begin{figure}[ht!]
  \centering
 %\fbox{\rule[-.5cm]{0cm}{4cm} \rule[-.5cm]{4cm}{0cm}}
  \includegraphics[width=5in]{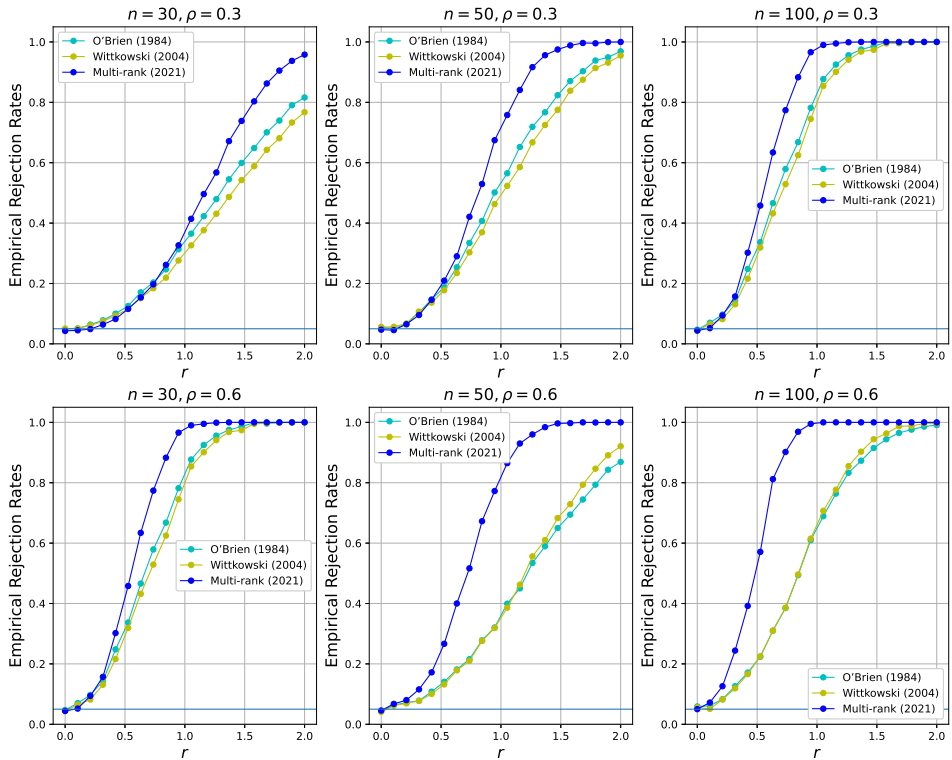}
  \caption{Empirical rejection rates for the simulation of time-to-event endpoint in Section \ref{sec:sim_survival}.} \label{figure:scenario_3}
\end{figure}

In this section, we carry out simulation studies to assess the finite sample performance of the test procedure on right-censored survival data. Let $S_{i}^x(t|\widecheck{\bm{x}}_i) = \mathbb{P}(T^x_{i}>t|\widecheck{\bm{x}}_i), S_{j}^y(t|\widecheck{\bm{y}}_j) = \mathbb{P}(T^y_{j}>t|\widecheck{\bm{y}}_j)$ be the survical function, representing the probability of surviving beyond time $t$, where $T^x_{i}, T^y_{j}$ are the survival times of subject $\bm{x}_i, \bm{y}_j$, respectively, and $\widecheck{\bm{x}}_i = (x_{i2}, \ldots, x_{id})^\top, \widecheck{\bm{y}}_j = (y_{j2}, \ldots, y_{jd})^\top$. We conciser the following Cox proportional hazards model
\begin{equation*}
S_{i}^x(t|\widecheck{\bm{x}}_i) = \exp\left(-H_0(t)\exp(\psi(\widecheck{\bm{x}}_i))\right), \,\,\, S_{j}^y(t|\widecheck{\bm{y}}_j) = \exp\left(-H_0(t)\exp(\psi(\widecheck{\bm{y}}_j))\right),
\end{equation*}
where $H_0(t)$ is the cumulative baseline hazard function and $\psi(\bm{x})$ is the covariates effect. We use the inverse probability method by \cite{bender2005generating} to generate $T_i^x, T_j^y$ from the hazard function. Specifically, let $U_i, U_j$ be uniformly distributed on $[0, 1]$, then the corresponding event time
\begin{equation*}\label{generate_T_from_hazard}
T_i^x = (S_{i}^x)^{-1}(U_i|\widecheck{\bm{x}}_i) = H_0(t)^{-1}\left(-\frac{\log(U_i)}{\exp(\psi(\widecheck{\bm{x}}_i))}\right), T_j^y = (S_{j}^y)^{-1}(U_j|\widecheck{\bm{y}}_j) = H_0(t)^{-1}\left(-\frac{\log(U_j)}{\exp(\psi(\widecheck{\bm{y}}_j))}\right).
\end{equation*}
In this simulation, we consider the number of endpoints $d=6$, where 
    \begin{align*}
    \widecheck{\bm{x}}_i = (x_{i2}, \ldots, x_{i6})^\top &\overset{\text{i.i.d.}}{\sim}\mathcal{N}(\bm{\mu}, \Sigma), \,\,\, \widecheck{\bm{y}}_j = (y_{j2}, \ldots, y_{j6})^\top \overset{\text{i.i.d.}}{\sim}\mathcal{N}(\bm{\mu} -r\bm{\nu}, \Sigma),
    \end{align*}
with $\bm{\mu} = (3, 2, 2, 1, 1)^\top, \,\,\, \bm{\nu} = (1, 0.1, 0, 0.1, 0.2)^\top$, $\sigma_{ij} = 1$ if $i=j$, and $\sigma_{ij} = \rho$ if $i\neq j$ representing the $(i,j)$th entry of the covariance matrix $\Sigma$. We consider $\rho = 0.3, 0.6$. In the context of the survival times $T_i^x, T_j^y$, we assume the baseline hazard function is
constant, i.e. the survival times are exponentially distributed which are generated from
\begin{equation*}
T_i^x = (S_{i}^x)^{-1}(U_i|\widecheck{\bm{x}}_i)=  -\frac{\log(U_i)}{\lambda\exp(\bm{\beta}^\top\widecheck{\bm{x}}_i)}, \,\,\, T_j^y = (S_{j}^y)^{-1}(U_j|\widecheck{\bm{y}}_j) = -\frac{\log(U_j)}{\lambda\exp(\bm{\beta}^\top\widecheck{\bm{y}}_j)},
\end{equation*}
with $\lambda = 0.1, \bm{\beta} = (0.5, 0.2, 0.3, 0.3, 0.5)^\top$. The corresponding survival endpoint $x_{i1}$ and $ y_{j1}$ are defined as $x_{i1} = \min\{T^x_{i}, C^x_{i}\}, y_{j1} = \min\{T^y_{j}, C^y_{j}\}$, where censoring times $C^x_{i}, C^y_{j}$ are generated from a uniform distribution $\bm{U}(0, 3)$ and  $\delta^x_{i}=\mathbbm{1}(T^x_{i} \leq C^x_i), \delta^y_{j}=\mathbbm{1}(T^y_{j} \leq C^y_j)$ denote the censoring indicator. To extend O'Brien's and Wittkowski's methods to survival endpoints, similar to the description in Section \ref{sec:survival}, we first use Wilcoxon pairwise comparison to obtain the relative rank of the survival time for each subject. We then replace the survival endpoint with the corresponding relative rank. We applied Algorithm \ref{alg:survival1} to examine the empirical size and empirical power of three testing procedures, and the results are summarized in Figure \ref{figure:scenario_3}. It can be observed that when $r=0$ (under $H_0$), the empirical rejection rates are all around $5\%$, indicating that the type I error is well controlled in all three testing procedures. However, as $r$ exceeds a threshold, the testing procedure based on the multivariate rank shows significantly better performance compared to the other two methods. Furthermore, as the correlation between each endpoint becomes stronger, the difference in power becomes larger, and Wittkowski's method outperforms O’Brien’s method when the correlation is stronger. These findings are consistent with the results observed in scenario 1. The results demonstrate the validity of Algorithm \ref{alg:survival1}.

\subsection{Sensitivity Analysis}
\begin{figure}[ht!]
  \centering
 %\fbox{\rule[-.5cm]{0cm}{4cm} \rule[-.5cm]{4cm}{0cm}}
  \includegraphics[width=5in]{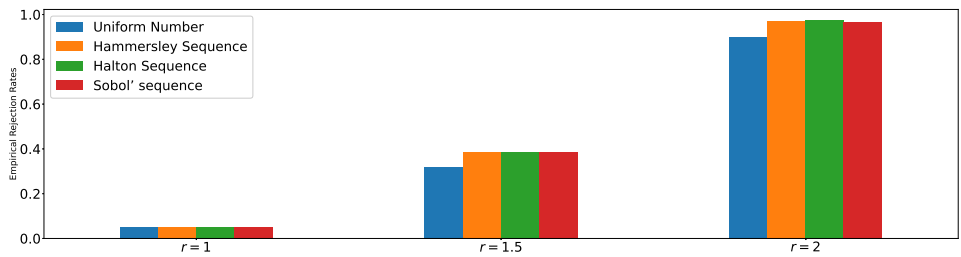}
  \caption{Empirical rejection rates of scenario 1 for various value of $r$ and $\rho=0.8$ using different low-discrepancy methods.} \label{figure:sensitity}
\end{figure}

%\section{Data Example} \label{Real_data}
In this section, we conduct a sensitivity analysis to examine the influence of the low-discrepancy sequence used in the construction of multivariate ranks, as described in Section \ref{Method}. We compare four different methods: uniform number in $[0, 1]^d$, Hammersley sequence \citep{hammersley1960monte}, Halton sequences \citep{Halton}, and Sobol' sequences \citep{Sobol}. For the uniform number, each component is generated from a standard uniform distribution. We evaluate the empirical rejection rates using scenario 1 with $\rho=0.8$. Figure \ref{figure:sensitity} presents the results, indicating that all methods effectively control the type I error. However, the low-discrepancy sequences (Hammersley, Halton, and Sobol') demonstrate higher power compared to the uniform numbers method. Importantly, the choice of low-discrepancy sequence does not significantly impact the performance. This finding is consistent with Figure \ref{figure:sobol_seq}, which illustrates the more even distribution of points provided by low-discrepancy sequences compared to randomly generated points.

\section{Conclusion} \label{Conclusion}
In this study, we studied a global nonparametric testing procedure based on multivariate rank for the analysis of multiple endpoints in clinical trials. We compared the multivariate rank approach with other two existing rank-based methods, namely O'Brien's rank-sum procedure and Wittkowski's method. Through extensive simulations, we observed that the multivariate rank approach consistently outperformed the classical methods in terms of both type I error control and power. The use of multivariate rank allowed us to directly incorporate the relationships among multiple endpoints in the testing procedure, providing a more comprehensive and informative analysis. This approach exhibited robustness against various data distributions and censoring mechanisms commonly encountered in clinical trials. Additionally, we conducted sensitivity analyses to assess the impact of low-discrepancy sequences on the performance of the multivariate rank-based approach. The results showed that incorporating low-discrepancy sequences, such as Hammersley, Halton, and Sobol', further enhanced the power of the method without compromising its overall performance. In conclusion, our study highlights the utility of the multivariate rank-based approach for the analysis of multiple endpoints in clinical trials. By leveraging the relationships among endpoints, this method offers improved power and robustness compared to existing rank-based methods. Further research could explore the extension of these methods to handle additional complexities and real-world clinical trial datasets.

\clearpage
\bibliographystyle{apalike}
\bibliography{ref_multirank_survival}

\end{document}